# Simultaneous induction of SSMVEP and SMR Using a Gaiting video stimulus: a novel hybrid brain-computer interface


Xin Zhang[1,2], Guanghua Xu[2*], Aravind Ravi[1], Sarah Pearce[3], and Ning Jiang[1*]

[1]Engineering Bionics Lab, the Department of Systems Design Engineering, University of Waterloo, Waterloo, ON, N2L3G1, Canada

[2]School of Mechanical Engineering, Xi'an Jiaotong University, Xi'an, Shaanxi, 710049, China

[3]Cognixion Inc., Toronto, ON, M5H 1K5, Canada



*Abstract*—**We proposed a novel visual stimulus for brain-computer interface. The stimulus is in the form gaiting sequence of a human. The hypothesis is that observing such a visual stimulus would simultaneously induce 1) steady-state motion visual evoked potential (SSMVEP) in the occipital area, similarly to an SSVEP stimulus; and 2) sensorimotor rhythm (SMR) in the primary sensorimotor area, because such action observation (AO) could activate the mirror neuron system. Canonical correlation analysis (CCA) was used to detect SSMVEP from occipital EEG, and event-related spectral perturbations (ERSP) were used to identify SMR in the EEG from the sensorimotor area. The results showed that the proposed visual gaiting stimulus-induced SSMVEP, with classification accuracies of 88.9 ± 12.0% in a four-class scenario. More importantly, it induced clear and sustained event-related desynchronization/synchronization (ERD/ERS) in the EEG from the sensorimotor area, while no ERD/ERS in the sensorimotor area could be observed when the other two SSVEP stimuli were used. Further, for participants with sufficiently clear SSMVEP pattern (classification accuracy > 85%), the ERD index values in mu-beta band induced by the proposed gaiting stimulus were statistically different from that of the other two types of stimulus. Therefore, a novel BCI based on the proposed stimulus has potential in neurorehabilitation applications because it simultaneously has the high accuracy of an SSMVEP (~90% accuracy in a four-class setup) and the ability to activate sensorimotor cortex. And such potential will be further explored in future studies.**

[†]his work was supported by National Natural Science Foundation of China (No. 51775415, and No.51505363), China Scholarship Council, and an NSERC-ENGAGE grant (No. 401261605). (*Corresponding authors: Guanghua Xu, Ning Jiang).






# I. INTRODUCTION

Motor impairments are the most common deficits following stroke (Schaechter 2004). Stroke patients could recover sensorimotor functions through rehabilitation training (Johansson 2000). During the rehabilitation process, it is important to provide proper feedback such that optimal patient engagement can be maintained. The feedback provided should be closely related to the user's own volitions (Burke et al. 2009). Thus, obtaining the volitions of the user is critical. However, in stroke rehabilitation, it is usually difficult, if not impossible, to obtain user's sensorimotor volition because of the patients' sensorimotor disabilities. For this reason, the brain-computer interface (BCI) is highly appealing as it is possible to obtain patients' volitions, such that their volitions can be explicitly utilized to provide proper feedback (Wolpaw et al. 2002). Another benefit of using BCI-based stroke rehabilitation is that when the detected sensorimotor volition of the patient is paired properly designed feedback, positive neural plasticity can be induced highly efficiently (Mrachacz-Kersting, Stevenson, et al. 2018b; Mrachacz-Kersting et al. 2016).

Currently, in the BCI literature, sensorimotor rhythm (SMR) (Ang and Guan 2017) and movement-related cortical potential (MRCP) (Mrachacz-Kersting, Stevenson, et al. 2018a) are two modalities of electroencephalograms (EEG) that have been investigated to detect motor imaginary (MI) in the context of stroke rehabilitation (Ang and Guan 2015; Mrachacz-Kersting, Jiang, et al. 2018). However, there are some critical limitations for these MI-based BCIs. First of all, a significant portion of the population (estimated 15% to 30%) cannot elicit sufficiently clear EEG pattern (Allison and Neuper 2010; Dickhaus et al. 2009). Further, MI-based BCIs using either SMR or MRCP still have limited performance. An averaged accuracy was 76.7% in detecting MI for walking from background idle state (a two-class scenario) (H. et al. 2014). Three classes were classified with an approximate accuracy of 70% in (LaFleur et al. 2013), and the average classification accuracy was 59.4% for a four-class scenario (Yao et al. 2018). Last but not least, recent research has shown that some patients experience full or partial loss of MI ability following stroke (Maciejewski, Withers, and Taylor 2013; Liepert et al. 2016; Ang et al. 2011). Only 68% of stroke patients operated the MI-based BCI better than chance level (Ang et al. 2012). As such, researchers started to explore alternative BCI paradigms with better accuracy for stroke rehabilitation. To this end, BCI based on steady-state visual evoked potential (SSVEP), which is an electrophysiological signal evoked by periodic visual stimulation (Zhang et al. 2015), is a potential candidate. Comparing with MI-based BCIs, SSVEP-based BCIs have the clear advantage of high decoding accuracy and calibration-free, or nearly calibration free (Kwak, Müller, and Lee 2015). SSVEP has also shown to exist in almost all subjects (Guger et al. 2012). The biggest concern of using SSVEP-based BCIs for stroke rehabilitation is that such stimuli activate the occipital area, and its effect on the sensorimotor cortex is unclear. Therefore, its relevance to stroke rehabilitation is in question.





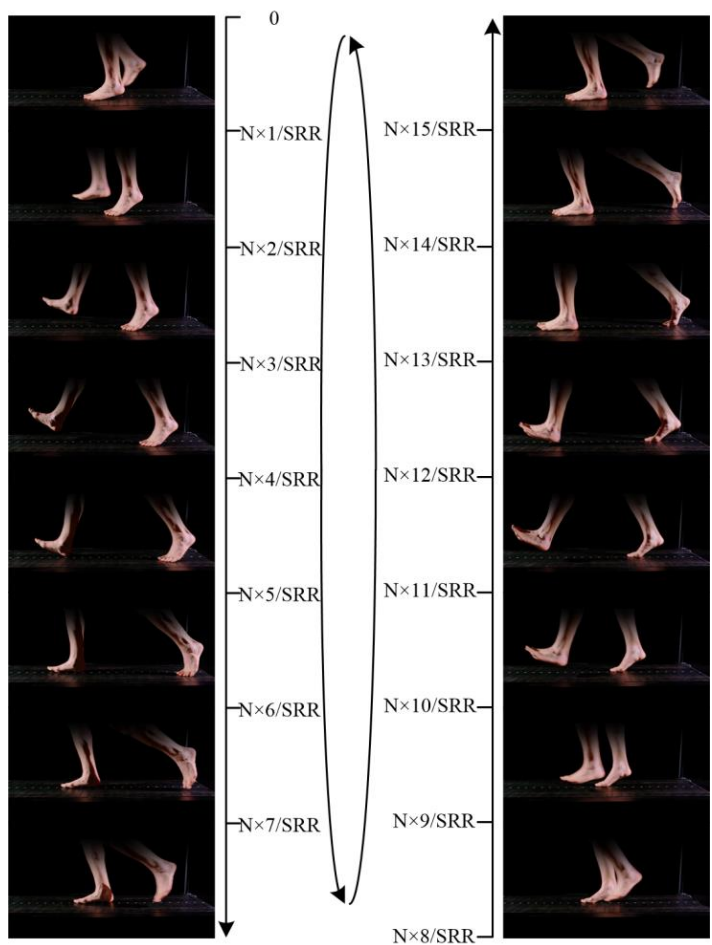

**Fig. 1. Generation of the gaiting stimulus.** *SRR* referred to screen refresh rate and *N* referred to the number of frames.

Therefore, for SSVEP-based BCI to have potential in stroke rehabilitation, one needs to at least demonstrate such BCI would be able to activate the sensorimotor area. For this purpose, action observation (AO) is a potential method that SSVEP-based BCI can leverage (Rizzolatti et al. 2014). AO has been applied in Stroke





(Yates, Kelemen, and Sik Lanyi 2016), Spinal cord injury (Collinger et al. 2014) and Parkinson Disease (Caligiore et al. 2017), where it has shown that AO can activate the motor neurons as those responsible for producing the observed action via the brain's mirror neuron system (MNS) (Tani et al. 2018; Cusack et al. 2016). Recently, it was shown that AO induced significantly stronger SMR than MI (Tani et al. 2018). And it was also suggested that changes in beta activity from sensorimotor cortex can be used to quantify activities of MNS (Muthukumaraswamy and Singh 2008), where it was suggested that AO could be a good option for patients with stroke who have difficulty using MI to effectively stimulate cortical-peripheral motor pathways. While most researchers focused on the response in sensorimotor cortex in AO, the response in occipital cortex seemed to be ignored.

Recently, Yan et al. (Yan et al. 2018) demonstrated that visual stimulus with periodic motion can induce steady-state motion visual evoked potential (SSMVEP), similar to SSVEP induced by traditional flashing stimuli. It showed that SSMVEP paradigms have low-adaptation characteristic and less visual discomfort for BCI applications. Indeed, many human actions are periodic motion such as gaiting. In AO, users watch similar human periodic movements. However, those human actions usually occur in frequencies much lower than the frequency range of SSVEP. As such, no research had attempted to investigate the possibilities of recognizing these low frequencies from occipital cortex.

In this study, we proposed a visual gaiting stimulus based on a video of human gaiting. We investigated the ability of this gaiting stimulus to 1) induce SSMVEP in the occipital cortex; 2) and to induce SMR in the sensorimotor area. Canonical correlation analysis (CCA) was used to recognize the four different stride frequencies from occipital EEG. And optimal CCA template signals were selected to improve the classification performance. Furthermore, event-related spectral perturbations (ERSP) in the sensorimotor cortex were used to quantify SMR. And these results were compared with the traditional SSVEP flicker stimuli and the checkerboard motion stimuli.





## II. METHODS

### A. Design of the visual gaiting stimulus

This study adopted the frame-based stimulation pattern to present the stimuli on a liquid crystal display (LCD) monitor. The frequencies of stimuli were determined by the frame rates. Images of one gait cycle were extracted from one video of human gait. A total of 16 captured images are used to generate the gaiting stimulus as shown in Figure 1. The screen refresh rate (SRR) was 60 Hz, *i.e.* 60 frames per second. For a visual stimulus, each of the 16 images would last $N$ frames, followed by the next image. Consequently, the frame rate of the stimulus was $F = \dfrac{60}{N}$ Hz. And the stride frequency, which is defined as the number of full gait cycle within 1 s, was $f = \dfrac{60}{16N}$. Thus, by using different $N$, stimuli with different frequencies of gaiting can be generated.

### B. Experiment protocol

Ten healthy subjects (20 to 30years old, 8 males and 2 females) participated in the experiments. The experimental protocol was approved by a University of Waterloo's Office of Research Ethics (# 23152), Canada. Written Informed Consent forms were obtained from the participants before their participation in the experiments.

Three types of visual stimuli (Flicker, Checkerboard, and Gaiting) were investigated in the current study. The program presenting the stimulus was developed with Matlab using the Psychophysics Toolbox (Brainard 1997). The first type of stimulus was the flicker stimulus. It is the traditional and most commonly used SSVEP stimulus (Cheng et al. 2002; Pan et al. 2013; Kwak, Müller, and Lee 2015). During the experiment, four flicker targets, as shown in Figure 2(A), were displayed on the screen with flicker frequencies of 8.57 Hz, 12 Hz, 10 Hz and 15 Hz in the left, right, up, and down position of the screen, respectively. The second type of stimulus was the checkerboard stimulus, a motion visual stimulus recently proposed (Zhang et al. 2017; Yan et al. 2018). The periodic expansion and contraction movements of the checkerboard could elicit SSMVEP in EEG from the occipital area when the participants gazed at the checkerboard (Yan et al. 2018).





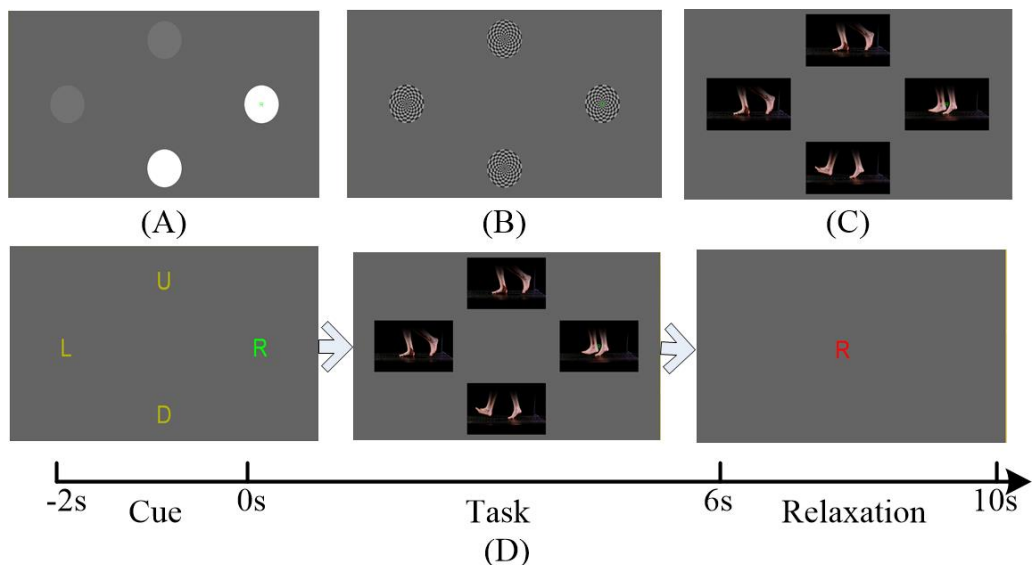

**Fig. 2. Illustration of the experiment protocol. (A) Flicker stimulus (Task 1). (B) Checkerboard stimulus (Task 2). (C) Gaiting stimulus (Task 3 & 4). (D) Illustration of one trial in the experimental runs, in which the Task was the gaiting stimulus.**

During the experiment, four checkerboard targets, as shown in Figure 2(B), were displayed on the screen with motion frequencies of $8.57\,\mathrm{Hz}$, $12\,\mathrm{Hz}$, $10\,\mathrm{Hz}$ and $15\,\mathrm{Hz}$ in the same position of the screen as the flicker targets. The third type of stimulus is the gaiting stimulus, described in Section II.A. Four values of the parameter $N$ were implemented: 7, 5, 6, and 4, resulting in the four gaiting stimuli placed in the same position of the screen as other two types of targets, as shown in Figure 2(C). The frame rates of these four stimuli were $8.57\,\mathrm{Hz}$, $12\,\mathrm{Hz}$, $10\,\mathrm{Hz}$ and $15\,\mathrm{Hz}$, consequently the stride frequencies were $0.536\,\mathrm{Hz}$, $0.75\,\mathrm{Hz}$, $0.625\,\mathrm{Hz}$ and $0.938\,\mathrm{Hz}$, respectively.

In one experimental session, the participant was seated in a comfortable chair and was briefed on the tasks to be performed. The participant was asked to watch the LCD screen on which the visual cues, stimuli, and feedback information were displayed (see Figure 2). The session consisted of a series of experimental runs, where a non-stop data recording was taken place. In each run, 20 experimental trials were performed by the participant. Each trial consisted of three phases: cue phase, stimulus phase and relaxation phase. Each trial started with the cue phase (from -2 to 0 seconds), where four cue letters ('L', 'R', 'U', 'D') would appear at the screen, at the left, right, up, and down positions. And one of the four letters would be green while the other three yellow (the left image of Figure 2(D)). The green letter indicated the target stimulus for the current trial, at which participant would then engage his or her gaze during the stimulus phase. The stimulus phase





would start at 0 second and last 6 seconds. In this phase, the four stimuli of one of the three types would replace the four letters, appearing on the screen for six seconds, during which the stimuli were modulated at the four frequencies stated above. The participants were asked to gaze at the target appearing at the same position as the green letter shown in the cue phase. The middle image of Figure 2(D) illustrates the stimulus phase of the gaiting stimulus. This was followed by the 4-second long relaxation phase, during which the participant could relax the gaze. At the same time, the online classification result using CCA would be displayed in the middle of the screen. Then the next trial would begin. The right image of Figure 2(D) shows an example of this phase, where the red letter 'R' indicates CCA classified that the participant was looking at the stimulus at the right side of the screen in the stimulus phase. In one experimental run, each of the four targets was repeated for five times, in random order. Each session contained four runs of each stimulus type, a total of 12 runs, also in random order. Task 1, Task 2, and Task 3 were used to denote the mental tasks the participant was asked to perform for the flicker, checkerboard and gait stimulus, respectively. Therefore, there were 80 trials for each stimulus type. The participants were asked to avoid moving their heads and avoid performing any sudden jerking movements during the experimental runs. A resting period was provided to the participants between runs to avoid fatigue. After twelve runs (four runs per stimulus type), a longer break was given. Then, two additional runs (Task 4) for the gaiting stimuli were performed. The difference between these additional runs from the earlier runs was that the participants were asked to imagine walking, mainly focus on imagining the movement in their feet, when they visually engaged at the gaiting stimuli targets.

*C. EEG acquisition*

EEG signals were recorded with a commercial research-grade EEG system (gUSBamp and Ladybird electrodes, g.tec Guger Technologies, Austria). Sixteen electrodes were placed at FCz, C1, Cz, C2, CPz, F3, F4, PO3, POz, PO4, PO7, PO8, O1, Oz, O2, and Iz of the international 10–20 system. Left earlobe was used as the reference and Fpz was used as ground. All electrodes' impedances were kept below $5\,\text{k}\Omega$ following the guideline provided by the manufacturer. The sampling frequency was $1200$ Hz. The signals were band-





**TABLE I The combinations of components in the CCA template signals**

| Combinations | COMPONENTS |
|---|---|
| $Cb1$ | $F$ |
| $Cb2$ | $F, F \pm 2 \times f$ |
| $Cb3$ | $F, F \pm 2 \times f, 2 \times F$ |
| $Cb4$ | *Select from* $F, F \pm 2 \times f, 2 \times F$ |

$F$ is the frame rate and $f$ is the stride frequency.

pass filtered between 0.1 and 100 Hz and a notch filter from $58$ Hz to $62$ Hz was used to eliminate the power

line interface. All EEG data and event time stamps (the beginning and end of each trial) were recorded for

subsequent processing.

*D. EEG analysis*

*1) The analysis of EEG from the occipital region*

The CCA algorithm is widely used in SSVEP processing, where it is used to calculate the correlations

between template signals and multi-channel EEG data (Zhonglin Lin et al. 2006). The formula of CCA is:

$$\text{Max}\left(\rho(x,y)\right) = \frac{E[w_x^T X Y^T w_y]}{\sqrt{E[w_x^T X X^T w_x] E[w_y^T Y Y^T w_y]}} \qquad (1)$$

where $\rho$ is the CCA correlation coefficient, $X$ is the template signals and $Y$ is the EEG data.

In this study, the EEG data $Y$ were composed of the EEG signals from channels PO3, POz, PO4, O1, Oz,

and O2. Data epochs were extracted according to recorded events. Considering a latency delay in the visual

system, a 140 ms delay was selected according to one prior study (Chen et al. 2015). Thus, the data epochs

were extracted from $0.14$s (time 0 represented when the stimuli targets occurred). Then the EEG data were

band-pass filtered from $4$ Hz to $50$ Hz. Furthermore, the templates $X$ were composed of several groups of

sine and cosine signals. The spectrum of the SSMVEP induced by the gaiting stimuli is more complex than

the other two types of stimuli as seen in Figure 3 (details in Section III.A). For this stimulus type, four

combinations ($Cb1$, $Cb2$, $Cb3$, and $Cb4$) of the frequency components were chosen as shown in Table 1. $Cb1$

was composed of the frame rate ($F$). $Cb2$ was composed of frame rate and the sum and difference between

$F$ and twice stride frequency ($f$), *i.e.* $F$ and $F \pm 2 \times f$. $Cb3$ included the second harmonic of $F$, *i.e.* $2 \times F$,

in addition to those in $Cb2$. And $Cb4$ was shown in (2).





$$Cb4 = \begin{bmatrix} F_1 & 2 \times F_1 & F_1 + 2 \times f_1 \\ F_2 & F_2 - 2 \times f_2 & F_2 + 2 \times f_2 \\ F_3 & F_3 - 2 \times f_3 & 2 \times F_3 \\ F_4 & F_4 - 2 \times f_4 & F_4 + 2 \times f_4 \end{bmatrix} \tag{2}$$

where $F_1 = 8.57$ Hz, $f_1 = 0.536$ Hz, $F_2 = 12$ Hz, $f_2 = 0.75$ Hz, $F_3 = 10$ Hz, $f_3 = 0.625$ Hz, $F_4 = 15$ Hz, $f_4 = 0.938$ Hz.

Thus, four kinds of templates $X$ were used for detecting SSMVEP in the gaiting stimuli in this study. And the $Cb4$ was used as the frequency components in the online analysis. When using $Cb4$ as the frequency components, the templates $X$ in CCA were shown in (3).

$$X = \begin{bmatrix} \sin(2\pi \times Cb4_{i,1} \times t) \\ \cos(2\pi \times Cb4_{i,1} \times t) \\ \sin(2\pi \times Cb4_{i,2} \times t) \\ \cos(2\pi \times Cb4_{i,2} \times t) \\ \sin(2\pi \times Cb4_{i,3} \times t) \\ \cos(2\pi \times Cb4_{i,3} \times t) \\ \sin(2\pi \times Cb4_{i,4} \times t) \\ \cos(2\pi \times Cb4_{i,4} \times t) \end{bmatrix}, i = 1, 2, 3, 4 \tag{3}$$

where $Cb4_{i,j}$ is the value in the row $i$ and in the column $j$ of $Cb_4$ in (2), $j = 1, 2, 3, 4$.

For the flicker and checkerboard stimuli, the template $X$ was composed of sine and cosine at the same frequency of the stimulus and its harmonics as shown in (4).

$$X = \begin{bmatrix} \sin(2\pi \times F_i \times t) \\ \cos(2\pi \times F_i \times t) \\ \sin(2\pi \times 2F_i \times t) \\ \cos(2\pi \times 2F_i \times t) \end{bmatrix}, i = 1, 2, 3, 4 \tag{4}$$

where $F_i$ is the frequency of each stimulus target. Note that because the SSVEP/SSMVEP spectrum from these two types of stimuli were simpler than that of the gaiting stimuli, therefore, the simpler CCA templates, with primary and second-order harmonics, was used for these two types of stimuli.

Besides, the target (all the three type stimuli in this study) on which the participant focused on could be identified by finding the maximal CCA coefficient.

*2) The analysis of EEG from the sensorimotor region*

To compare the different effects of the three stimuli and motor imagery on the sensorimotor areas, the





analysis of EEG data from the sensorimotor region were performed using EEGLAB (Delorme and Makeig 2004). First, the EEG data were filtered from $4$ Hz to $50$ Hz. Then the EEG data were visually inspected for the artifacts (e.g. electrode cable movements, swallowing, etc.) and affected trials were removed from further analysis. On average, $81.19 \pm 5.33\%$ (mean + standard deviation) of the trials of each participant's EEG data remained in subsequent analyses.

Next, the preprocessed datasets containing EEG and electrooculogram (EOG) were decomposed by independent component analysis (ICA) (P.~Common 1994). ICA was performed on individual subjects over all trials within one type of stimulus. Based on the individual component scalp maps and component activations (scroll), the component mainly contained EOG was rejected. As we were interested in lower extremity motor functions, we focused on the channel Cz. Therefore, Laplacian spatial filtering was applied to Cz, and four channels surrounding it: FCz, C1, C2 and CPz. Laplacian filtering has shown to improve quality of sensory-motor rhythm estimation (Kayser and Tenke 2015).

Finally, the event-related spectral perturbation (ERSP) (Makeig 1993) were computed based on the EEG data from channel Cz. Relative changes in spectral power were obtained by averaging the difference between each single-trial log spectrogram and baseline $(-1.9\text{–}0\text{ s})$. The ERD index was calculated as the log ratio of the power in a certain frequency band during each condition and the power during the baseline, which was from $-1.9\text{–}0$ s. The ERD index was calculated (Lim and Ku 2018):

$$\text{ERD Index} = 10 \times \log_{10} \frac{P}{R} \tag{5}$$

where $R$ is the power in the reference period (baseline) and $P$ is the power during the task period.

This index quantifies the degree of EEG band-power reduction resulting from the desynchronization of cortical neurons when executing a motor task (Li et al. 2015). And the frequency band mu-beta $[8, 26]$ Hz were adopted in this study.

### E. Statistical analysis

A mixed-effect analysis of variance (ANOVA) was used for the analysis of the classification accuracies of the gaiting stimuli using different CCA templates. The frequency combination was a fixed factor with four levels: $Cb1$, $Cb2$, $Cb3$, and $Cb4$. And the participant was a random factor. Similarly, mixed-effect ANOVA was used to analyze the ERD index. Task was a fixed factor with four levels: 1) observing the flicker stimuli; 2) observing the checkerboard stimuli; 3) observing the gaiting stimuli, and 4) simultaneously observing the gaiting stimuli and performing motor imagery. Group was another fixed factor with two levels based on the classification accuracy (see Section IIIA for details about the grouping). The subject was a random factor. The Bonferroni post hoc analysis was used to test the significance. And the statistical significance level was 0.05 for all tests.





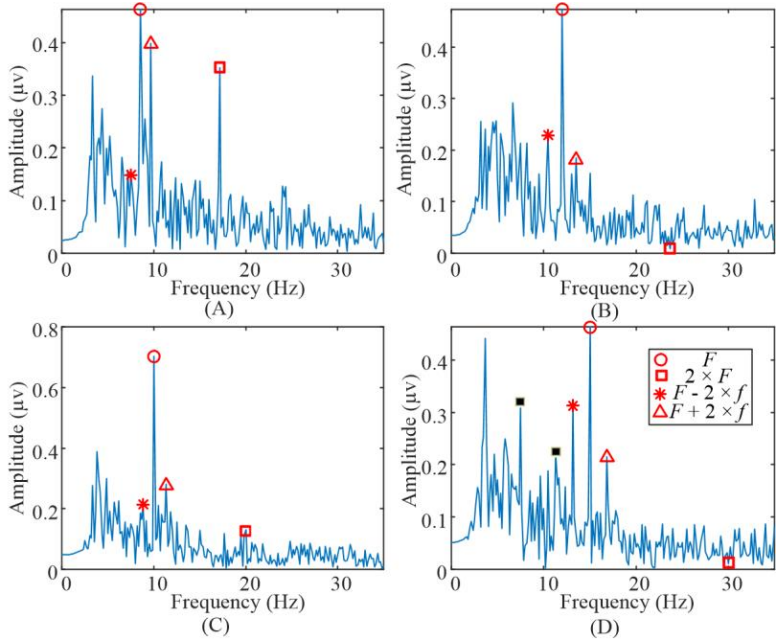

**Fig. 3.** Spectrums of the EEG data from channel Oz for the gaiting stimulus with different frame rates ($F$) and stride frequencies ($f$). (A) $F = 8.57$ Hz, $f = 0.536$ Hz. (B) $F = 12$ Hz, $f = 0.75$ Hz. (C) $F = 10$ Hz, $f = 0.625$ Hz. (D) $F = 15$ Hz, $f = 0.938$ Hz. The two black squares were at 7.5 Hz and 11.25 Hz, which is coincidentally $F$-$2\times f$ in (A) and $F$+$2\times f$ in (C). Detail discussion on this in Section III.A.

## III. RESULTS

### A. Analysis of occipital EEG

For each gaiting stimulus target, the spectrum of the EEG signals from Oz electrode was calculated by averaging across all the trials from all participants. The entire 6s EEG data, when participants were observing the stimuli, were used to calculate the spectrum (the frequency interval is 0.167 Hz). Figure 3 showed the mean amplitude spectrum of each gaiting stimulus target. The peaks in the spectrum could be clearly





identified exactly at the frame rate, the sum and the difference between the frame rate and twice the stride frequency. Additionally, the peak at the second harmonic of the frame rate could also be found in Figure 3(A). Evidently, the proposed gaiting stimulus can evoke SSMVEP in the occipital region, although the elicited frequency components in EEG are more complex than the regular flicker and checkerboard stimuli. And we could choose the temple signals for CCA from the combinations of these frequency components (see Table I).

Table 2 showed the CCA accuracies (mean ± standard deviation) in gaiting stimuli using different combinations in template signals (see Section II.D.1). The window length of the EEG data was 6 s. According to the ANOVA analysis, the factor Frequency Combination had a significant effect on the accuracies (F(3,27)=7.48, $p$=0.001). The post-hoc comparison revealed that the accuracies using Cb4 were significantly higher than the accuracies using $Cb3$ ($p$=0.002) and $Cb2$ ($p$=0.004). While there was no significant difference between the accuracies using $Cb1$ and $Cb2$ ($p$=0.232), $Cb1$ and $Cb3$ ($p$=0.149), $Cb1$ and $Cb4$ ($p$=0.606).

To further compare the relative classification performance of the gaiting stimulus targets in different combinations in CCA template signals, the confusion matrices of the classification accuracies were calculated as shown in Figure 4. We observed that target 4 resulted in the lowest classification accuracy in all cases and it was most frequently misclassified as target 1 and target 3 when using $Cb2$ and $Cb3$. This poor performance is likely due to some confounding frequency components for these targets. Because $F_1 = 2 \times f_1$ (7.5 Hz) and $F_3 + 2 \times f_3$ (11.25 Hz) are the coincided with the frequency components of the spectrum of target 4: $8 \times f_4$ and $12 \times f_4$. These confounding frequency components were shown as the two black squares in Figure 3(D). Because $F_1 - 2 \times f_1$ and $F_3 + 2 \times f_3$ were not selected in $Cb4$, the correct classification accuracies in target 4 increased approximately 10%, 30% and 30% comparing with $Cb1$, $Cb2$, and $Cb3$, respectively. This could be the same reason for the poorer performance when using $Cb2$ than using $Cb1$ for target 2. Therefore, the modulation frequency components in SSMVEP should be carefully considered in designing such a paradigm and data analysis, i.e. some calibration should be required. In the current study, $Cb4$ was used for the gaiting stimulus.

**TABLE 2 Accuracies of the gaiting stimuli under different template signal combinations**

| Subjects | Accuracy (%) | | | |
|---|---|---|---|---|
| | Cb1 | Cb2 | Cb3 | Cb4 |
| S1 | 97.50 | 97.50 | 97.50 | 97.50 |
| S2 | 93.75 | 100.00 | 100.00 | 100.00 |
| S3 | 98.75 | 82.50 | 82.50 | 97.50 |
| S4 | 65.00 | 66.25 | 66.25 | 78.75 |
| S5 | 73.75 | 66.25 | 67.50 | 77.50 |
| S6 | 96.25 | 97.50 | 96.25 | 97.50 |
| S7 | 95.00 | 93.75 | 92.50 | 96.25 |
| S8 | 95.00 | 81.25 | 81.25 | 96.25 |
| S9 | 83.75 | 73.75 | 72.50 | 82.50 |
| S10 | 58.75 | 58.75 | 57.50 | 65.00 |
| mean ± std | 85.75 ± 14.75 | 81.75 ± 15.08 | 81.38 ± 14.99 | 88.88 ± 12.01 |





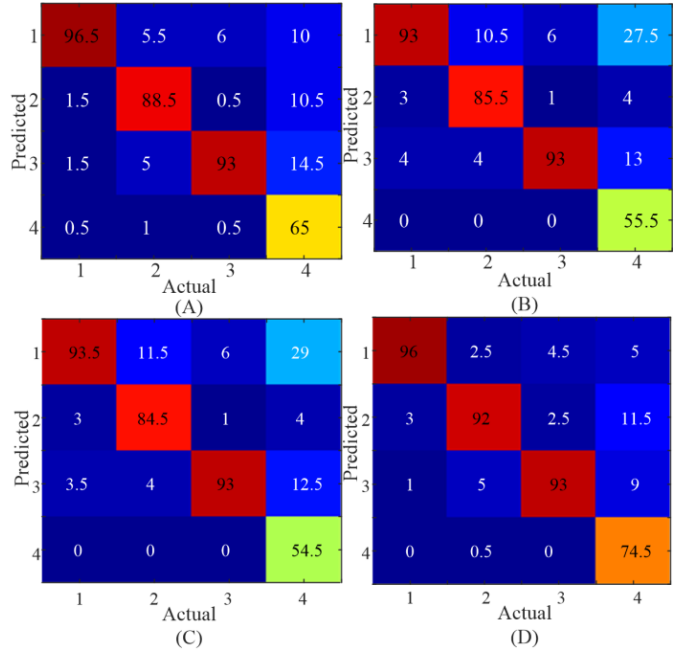

**Fig. 4.** **Confusion matrices for the gaiting stimulus using different combination components in the template signals of CCA to do classification. The color scale reveals the average classification accuracies (%), the diagonals labeled with the correct classification accuracies among all the participants (%). (A) using Cb1. (B) using Cb2. (C) using Cb3. (D) using Cb4.**

Finally, to compare the classification performance for the three types of stimulus, the average accuracies (mean $\pm$ standard deviation) with different lengths of the processing window (from 1 s to 6 s with a step of 1 s) were calculated and showed in Figure 5. It was observed that, in general, the average accuracies increased with longer time window lengths, regardless of the type of stimulus. For the flicker stimulus, the accuracy was $89.5 \pm 8.60\%$ when the processing window was 1 s long. The accuracy further increased for longer processing window lengths, reaching a stable and high accuracy level at 2 s and longer. These results were similar to values reported in the literature (Suefusa and Tanaka 2018; Zhonglin Lin et al. 2006; Chen et al. 2015) for such stimulus. For the checkerboard stimulus, the performance was in general similar to that of the flicker stimulus, albeit with a lower accuracy. The performance level was also similar to previous studies (Zhang et al. 2017). For the gaiting stimulus, the accuracy at 1 s was only $34.8 \pm 8.78\%$, significantly lower than the other two types. Its performance progressive increased with longer window lengths, reaching $88.9 \pm 12.0\%$ at 6-s window. In general, it also had larger variability at each windows length than the other two types of stimulus. The result indicated that the gaiting stimulus needed longer stimulus duration than the other two types of stimulus to reach a stable and high accuracy level.





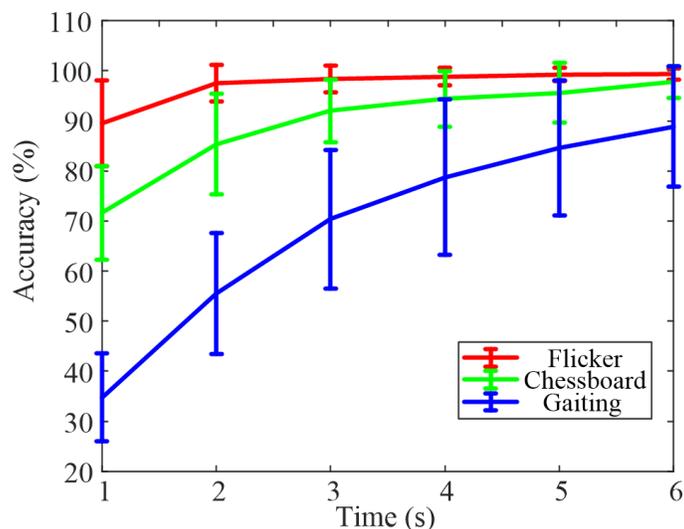

**Fig. 5. Classification accuracies with different time window lengths for the three kinds of stimulus.**

### B. Analysis of sensorimotor EEG

Figure 6 presents the grand average ERSP from channel Cz across the data from all participants. Evidently, during the task period (from 0 to 6 s), both observing the gaiting stimulus (Figure 6(C)) and imagery (Figure 6(D)) by the participants evoked clear and sustained ERD, which was followed by a clear ERS during relaxation period (after 6 s). Both the ERD and ERS in the beta band. As well, the desynchronization in motor imagery was visibly stronger than in the observation of gaiting stimulus. In comparison, observing the flicker and checkerboard stimuli did not show visible SMR at all.

To further investigate the effects of the four tasks on the sensorimotor area EEG, the ERD indexes in the mu-beta band during the task period were calculated and presented in Figure 7. We separated the participants into two groups (Group 1 and Group 2) based on the classification accuracies using Cb4 in Table 2. Group 1 consisted of participants 1, 2, 3, 6, 7, and 8, whose accuracies were higher than 85%. Other participants were in Group 2. We used a mixed effect model of ANOVA to quantify the differences. The tasks had a significant effect on ERD index values (F(3,24)=4.13, $p$=0.017). Based on the Bonferroni Pairwise comparison, there was no significant difference in ERD index values between task 3 and task 4 ($p$=1) in Group 1. The ERD index values in task 3 were significantly lower than the values in task 1 ($p$=0.012) and 2 ($p$=0.04) in Group 1. While there was no significant difference in the ERD index values between task 3 and 1 ($p$=1), task 3 and 2 ($p$=1) in group 2.The result indicated that mainly those participants with sufficiently clear SSMVEP pattern (classification accuracy > 85%) had strong ERD in the mu-beta band when observing the gaiting stimuli.





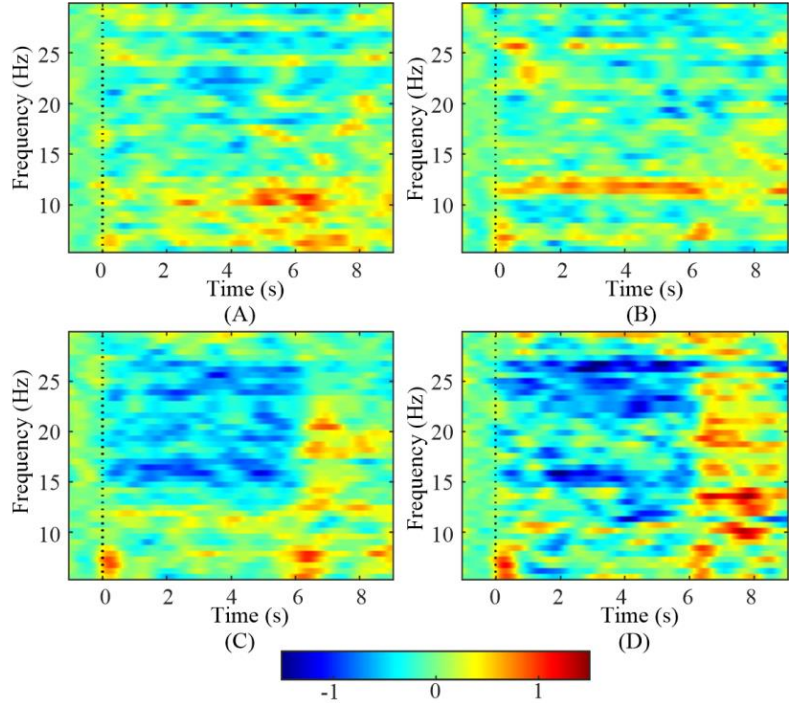

**Fig. 6. Grand average ERSP from channel Cz and the ERSP images show the time-frequency results during different tasks: observing different type of stimulus and imagery. (A) Observe the flicker stimulus. (B) Observe the checkerboard stimulus. (C) Observe the gaiting stimulus. (D) Observe gaiting stimulus and motor imagery. Time 0 s indicates beginning of the task and time 6 s indicates the completion of the task.**

## IV. DISCUSSIONS AND CONCLUSION

In this study, we proposed a visual gaiting stimulus that is capable of simultaneously inducing SSMVEP from the occipital area and SMR from the sensorimotor area. We compared the proposed gaiting stimulus with two other SSVEP/SSMVEP stimuli: flicker and checkerboard. We showed that the main SSMVEP frequencies induced by the proposed gaiting stimulus were at the frame rate and the sum and difference of the frame rate and twice stride frequency. And the four-class classification accuracy reached $88.9 \pm 12.0\%$ using CCA. More importantly, the proposed gaiting stimulus induce clear ERD/ERS in beta band in the sensorimotor area (Figure 6(C)), while no clear ERD/ERS could be observed when flicker and checkerboard stimuli were used (Figure 6(A) and 6(B)). Therefore, a novel BCI paradigm, which based on this gaiting stimulus, simultaneously enjoys 1) the benefit of high multi-class decoding accuracy of SSVEP paradigm; and 2) the ability to activate sensorimotor area similarly to MI-based paradigms.





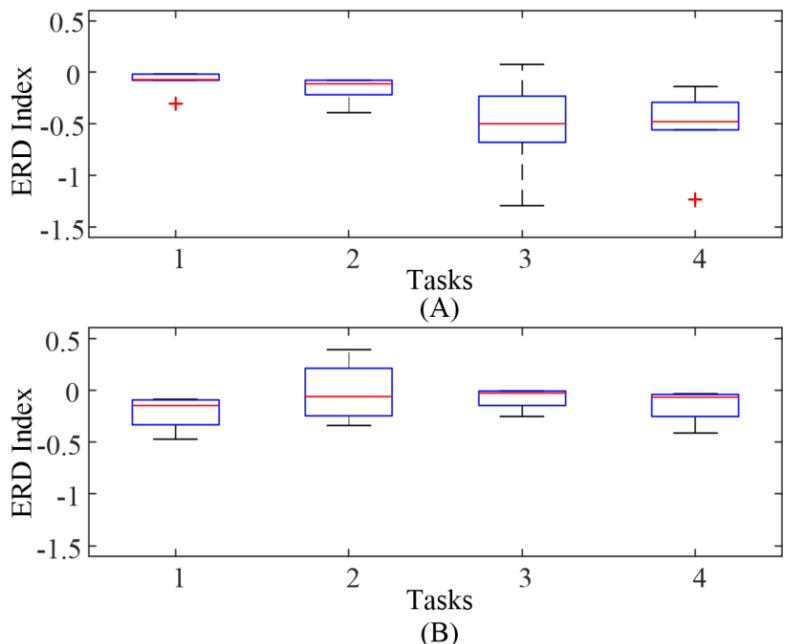

**Fig. 7. The ERD index in mu-beta band in different tasks (1: observe the flicker stimuli, 2: observe the checkerboard stimuli, 3: observe the gaiting stimuli, 4: observe the gaiting stimuli and motor imagery). (A) The ERD index in Group 1. (B) The ERD index in Group 2.**

The underlying nonlinearity of the EEG response to stimuli using dual frequencies was first reported by Regan et al. (Regan and Regan 1988). Several follow-up studies, using the simultaneous modulation of stimulus luminance, reported similar results ($m \times f \pm n \times f$) (Srihari Mukesh, Jaganathan, and Reddy 2006; Shyu et al. 2010; Chang et al. 2014). In our previous study, we explored the brain response to the motion-based stimulation with equal luminance using inter-modulation frequencies (Zhang et al. 2017). And the results also showed the underlying nonlinearity of the EEG response to motion stimulus. The objective of these previous studies was increasing the number of targets with limited stimulus frequencies, consequently





increasing the information transfer rate of the BCI. In the current study, we used the characteristics of the modulation stimulus to realize the recognition of stride frequency, which was not in the regular frequency regions in SSVEP. And we demonstrated that the stride frequencies could be identified by the modulation on frame rate. The main peaks in the spectrum were at the frame rate, the sum and the difference of frame rate and twice stride frequency. Our recent study explored the characteristics of brain response to a light spot humanoid motion stimulus modulated by the change of brightness, where the main peaks were at $F$, $F - f$, and $F + f$ ($F$: the high flicker frequency, $f$: stride frequency) (Zhang et al. 2018). It is possible that the participants were difficult to distinguish the left and right feet in this study. Thus, the twice stride frequency modulated onto the side band of frame rate. Further, the SSMVEP responses induced by the gaiting stimulus were mainly caused by the periodic motion. Thus, other types of human periodic motion, generating in the same method as the gaiting stimulus, could induce similar SSMVEP features.

As shown in our analysis, to detect SSMVEP induced by the proposed gaiting stimulus, the template signals for CCA needed to be carefully selected. All the frequencies in $m \times f \pm n \times f$ could be induced by the stimulus with modulation and should be examined for possible inclusion into the templates. The reasons why these frequency components occurred in the SSMVEP were still unknown and needed to be explored in the future. However, one rule to select the template signals was avoiding using the same frequency components across multiple stimuli, as shown in this study (the two black squares in Figure 3(D)). After properly selecting the template signals, the classification accuracies significantly increased.

To our best knowledge, this is the first study exploring the responses of the SSVEP/SSMVEP paradigms in the sensorimotor cortex. The results showed that the SSVEMP/SSVEP classification accuracy of the proposed gaiting stimulus is similar to that of the other two conventional stimuli. On the other hand, we showed that the proposed stimulus activates the sensorimotor area in a similar fashion as MI-based BCIs, but the traditional SSVEP flicker stimuli and the checkerboard motion stimuli do not have this capability. Mu and beta suppression have been widely used to explore the MNS, while some researchers still concerned that changes in the mu power may be driven largely by attentional processes rather than mirror neuron activity





(Aleksandrov and Tugin 2012; Hobson and Bishop 2016). Also, we can find ERS in Fig. 6 when observing the stimulus. That might be the stimulus frequencies. Previous studies also showed that SSVEP response is widely distributed over the occipital and the other areas, including parietal and frontal lobes (Srinivasan, Bibi, and Nunez 2006; Sperling, Ding, and Srinivasan 2005). Thus, we combined the mu band and the beta band together to calculate the ERD index to compare the different effects on the sensorimotor cortex with different stimuli.

Our results indicated that the proposed gaiting stimulus could not activate the sensorimotor cortex for every participant sufficiently. The gaiting stimuli mainly activate the participants with good performance of classification in SSMVEP of the gaiting stimuli. That was reasonable as poor performance of classification in SSMVEP meant the occipital cortex received poor information about the gaiting stimuli, which could due to poor engagement of the participant. This correlation indicated that the accuracies in the gaiting stimuli among different participants might be used to assess the feasibility of individuals to use the proposed BCI.

Recently, Ku, et al. (Lim and Ku 2017, 2018) used the flickering action video as the stimulus to induce SSVEP and produce MNS activation. And they used the SSVEP response to classify whether the stimuli were being attended to (Lim and Ku 2018). However, in the setup of this study by Ku et al., it was not possible to identify whether the participants engaged in the action or the background in the video. If the participant engaged at the background of the video (flickered white and black), they would still get the SSVEP response, without activating the sensorimotor area based, on the results in this study. In our study, it was the motion of gait, not the flicker in the video, that induced SSMVEP. And the source of inducing SSMVEP and activating the sensorimotor cortex was the same. This might be also the reason for the relationship between the decoding accuracies and level of ERD.

This study compared three types of SSVEP/SSMVEP stimuli and the results showed that the proposed gaiting stimulus can simultaneously activate the sensorimotor cortex and induce SSMVEP. Thus, a BCI based on this type of gaiting stimuli combines with the benefit of SSVEP-based BCI (high accuracy and calibration-





free) and the ability of MI-based BCIs in activating sensorimotor area. Such a BCI paradigm is appealing to and has potential in neurorehabilitation applications. However, in order to further explore such potential, one needs to demonstrate the ability of the proposed BCI in inducing cortical plasticity. A randomized controlled study, where the effects of such BCI on neural plasticity is measured by the standard trans-cranial magnetic stimulation (TMS), should be conducted. Further, combining SMR features and SSMVEP features might have the ability to improve the performance of detecting brain switch in AO, and should be investigated in our next studies.

## Acknowledgment

We want to thank the participants for participating in these experiments and anonymous reviewers for their helpful comments.